
\def \ltsima{$\; \buildrel < \over \sim \;$}
\def \simlt{\lower.5ex\hbox{\ltsima}}            
\def \gtsima{$\; \buildrel > \over \sim \;$}
\def \simgt{\lower.5ex\hbox{\gtsima}}            
\def \rat{t_c / t_d}
\magnification=1200
\vsize=23 true cm
\centerline {\bf A MODEL FOR THE X-RAY AND UV EMISSION FROM SEYFERT GALAXIES}
\centerline{\bf AND GALACTIC BLACK HOLES}
\vskip 0.5 true cm
\centerline{Francesco Haardt$^{1,2}$, Laura Maraschi$^3$
and Gabriele Ghisellini$^4$}
\vskip 0.4 true cm
\item{1:}
ISAS/SISSA, Via Beirut 2--4, 34014 Trieste, Italy, E-mail: 38028::haardt\par
\item{2:} present address: STScI, 3700 San Martin Dr.,
Baltimore, MD 21218, E-mail: 6559::haardt \par
\item{3:} Universit\'a di Genova, Genova, Italy, E-mail: 32366::maraschi\par
\item{4:} Osservatorio di Torino, Pino Torinese, 10025 Torino, Italy, E-mail:
32065::ghisellini

\vskip 1 true cm

\centerline{\bf ABSTRACT}
\vskip 0.5 truecm
We propose that the X--ray emission from radio quiet AGN and galactic black
holes is due to Comptonization of soft thermal photons emitted by the
underlying accretion disk in localized structures (blobs). The power per unit
area produced by the blobs, impinging on the disk, can easily dominate the
radiation internally produced by the disk.
In this case the electron
temperature and the high energy spectrum can be determined in a similar way as
in the previously studied homogeneous model (Haardt \& Maraschi 1991). However
in the present model: a) the emitted spectrum is largely independent of the
{\it fraction} of gravitational power dissipated in the blobs; b)
the X--ray spectrum can be harder depending on a form factor of the blobs; c)
the UV (or soft X--ray for galactic objects) luminosity that is
not intercepted by the blobs can be larger than the X--ray luminosity.
In the framework of a simplified
accretion disk $\alpha-\Omega$ dynamo model, we make order of magnitude
estimates of the number of active blobs, their size, luminosity and hence their
compactness, finding values in agreement with what is observed. The expected UV
to X--ray spectra and correlations of  X--ray  and UV light  curves
are discussed.
\vskip 0.3 truecm
\noindent
{\it Subject headings:}
Accretion disks --
plasmas --
radiation mechanism: thermal --
galaxies: Seyferts --
X--rays: general
\vskip 0.3 truecm
{\centerline{\it to be published in Ap.J. Letters}}

\vfill
\eject

\centerline{\bf 1. INTRODUCTION}
\vskip 0.5 truecm

\noindent
The recent observations by OSSE show spectral ``breaks" or cut--offs
in the hard X--ray spectra of some Seyfert 1 galaxies (e.g. Maisack et
al. 1993, Cameron et al. 1993), supporting thermal or quasi--thermal models
for the X--ray emission, in which a population of semi--relativistic
electrons Comptonizes the available soft photons (e.g. Shapiro, Lightman \&
Eardly 1976, Podzniakov, Sobol \& Sunyaev 1980, Sunayev \& Titarchuk 1980),
and reinforcing the analogy with the high energy spectra of galactic
black hole candidates (GBH) (e.g. Done et al. 1990, Johnson et al. 1993,
and Grebenev et al. 1993).

A second implication of the OSSE observations is that the luminosity
$L_X$ in the medium and hard X--ray range can be reliably estimated.
If the ``X--ray bolometric correction" established for the few galaxies
detected by OSSE is adopted as a general property, the X--ray to UV flux
ratios measured with ROSAT for a large sample of objects (Walter \& Fink
1993) leads to an estimate of the X--ray luminosity in many cases
definitely smaller than that at the UV bump, $L_{UV}$.
On the other hand in the best studied objects, the Seyfert galaxies NGC
4151 and NGC 5548, the UV and X--ray fluxes are comparable, and vary in
a correlated fashion on time-scales of weeks (Perola et al. 1989, Clavel
et al. 1992, hereinafter C92).
This led to the suggestion that in these objects the UV emission is
due to reprocessing of the higher frequency radiation, implying
$L_{X}\simgt L_{UV}$.

We have proposed (Haardt \& Maraschi 1991, Haardt \& Maraschi 1993,
hereinafter Papers I and II respectively) that the main features of the
high energy emission from radio quiet active galactic nuclei (AGN), and
GBH (Haardt et al. 1993, Ueda, Ebisawa \& Done 1994),
can be explained by  the interplay
of a hot active corona with a colder accretion flow.
Soft thermal photons with an energy of few tens (hundreds for GBH) eV
are Comptonized by mildly--relativistic electrons in the hot corona,
leading to the formation of a power--law spectrum with a high energy
cut--off.
In steady state, the equilibrium temperature of the electron
distribution (assumed to be Maxwellian) can be computed balancing
heating and radiative cooling, and depends only on the electron
scattering optical depth $\tau$.
Furthermore, when the compactness $\ell$ of the source
(proportional to the luminosity to size ratio) is large,
electron--positron pair production  yields a lower limit to
$\tau$ and an upper limit to the temperature $kT$.
Quite remarkably, for $10\simlt \ell \simlt 100$ the theoretical values
$300\simgt kT \simgt 50$ keV are close to the first results of the OSSE
experiment (e.g. Maisack et al. 1993, Cameron et al. 1993).

The average observed X--ray spectrum can be reproduced if the
Comptonized--to--soft luminosity ratio $L_C/L_S$ is $\simeq 2$.
In fact this value leads to a Compton $y$ parameter close to 1, and to a
spectral index of the Comptonized spectrum $\alpha_x\simeq 1$.
In Paper I and II the condition $L_C/L_S \simeq 2$ was achieved by
assuming that the {\it entire} gravitational power is released
in the hot corona.
Therefore the soft emission $L_S$ derived only from absorption and
reprocessing of the high energy flux impinging on the disk.
In that model, roughly half of the Comptonized photons is absorbed in
the cold disk
and reemitted as  thermal radiation, while half is radiated away.
Any further local dissipation within the disk would produce additional
soft photons, lowering the coronal temperature and hence giving rise to
steeper X--ray spectra.
The model is then tightly constrained, and predicts nearly equal UV and
X--ray luminosities. As was illustrated in Paper II, the different angular
distributions of the Comptonized photons with respect to the thermal ones
can give rise to a UV flux larger than the X-ray flux for viewing angles
close to face on, but it seems difficult to achieve
soft--to--hard ratios greater than 5. Another prediction
is a tight correlation between the hard
Comptonized photons (X--rays) and the soft photons emitted by
the cold disk (UV or soft X--rays for GBH).
For NGC 4151 and NGC 5548, the UV and  X--ray fluxes simultaneously
observed in different periods are indeed correlated, but  variations (up
to 30 \%, Nandra et al. 1990) in X--rays on timescales of hours are not
accompanied by similar variations in the UV.
Furthermore, for large UV fluxes, the correlation between X--rays and UV
breaks down (Perola et al. 1986, C92).

In Papers I and II we considered a uniform, plane parallel model
which allowed us to minimize the parameters of the problem.
Here we wish to relax these assumptions and adopt a more physical
description of the energy dissipation process in order to account for
a broader set of observational results, including X--ray and UV
variability.

We maintain the assumption that the disk magnetic field can drain a
fraction $f$ of the accretion power outside the accreting flow, but
assume that this power is dissipated in active localized blobs which
cover a small fraction of the total disk area.
In \S 2 we discuss the effects of the new assumptions on the observed
emission spectral shape.
In \S 3 we outline a plausible scenario that can lead to the formation
of a structured corona and we discuss a specific model following
Galeev, Rosner \& Vaiana (1979, hereafter GRV).
We also estimate the number of active blobs,
their size, luminosity and hence their compactness.
Finally, in \S 4 we discuss our results and in \S 5 we present a brief
summary of our results.

\vskip 1 true cm
\centerline{\bf 2. STRUCTURED CORONA: PHENOMENOLOGICAL APPROACH}
\vskip 0.5 true cm
\noindent

We assume that the accretion disk dissipates internally a fraction $(1-f)$ of
the accretion power $Q$ [erg cm$^{-2}$ s$^{-1}$], while the remaining fraction
$f$ is stored in magnetic field structures which lead to active blobs of
typical size $R_b$.
We also assume that the energy is stored in the magnetic
field in a `charge time' $t_c$ but is released on a much shorter
`discharge time' $t_d$.
The ratio $t_d/t_c$ clearly coincides with the fraction of the disk area
which is covered by active blobs at any time.
The luminosity of a single blob will be
$$
L_{blob}\, = \, [f\, Q\, \, {t_c \over t_d}]\,\pi\,R_b^2.
\eqno(1)
$$
Assuming that roughly half of the blob luminosity is intercepted and
reprocessed locally by the underlying disk, the total luminosity
crossing the blob is
$$
L_{cool}\,=\,[Q_{disk}\,+\, Q_{rep}]\,\pi\,R_b^2 =\,
\left[(1-f)Q\,+\,0.5\, C \, f \, Q \, {t_c \over t_d}
\right]\,\pi\,R_b^2.
\eqno(2)
$$
The parameter $C$ indicates the fraction of reprocessed radiation
which crosses the blob.
In the plane parallel limit $C=1$ as the whole reprocessed flux is
effective in cooling the corona, while a height--to--radius ratio for the
blobs of order one leads to $C\simeq 0.5$.

Writing the Compton amplification factor as $(A-1)=L_{blob}/L_{cool}$,
we see that $A$ is a function of three parameters, namely $f$, $C$
and $t_c/t_d$. However in the limit $f\gg t_d/t_c$ we have
$$
(A-1)\simeq 2/C, \eqno (3)
$$
and only $C$ plays a role in determining the emitted spectrum.
This condition means that the disk emission per unit area
below the blobs is dominated by the reprocessed radiation
originally emitted by the blobs themselves.
If $t_d/ t_c$ is small, we can obtain the `right'
$\alpha_x$ nearly independently of the value of $f$: large UV
luminosities are thus possible even if the reprocessed radiation
dominates the cooling of the blob.
As the ratio of the disk to the {\it observed} blob emission is
$2(1-f)/f$, in principle $f$ can be derived from observations.

The actual value of $\alpha_x$ is determined by $C$.
In the geometrically thin case we have $C=1$ and we obtain the
results discussed in Papers I \& II.
If the blobs have a height--to--radius ratio of the order of one or more,
$C<1$,
the source is {\it photon starved}, and the corresponding power law spectral
index is flatter.
The observed  spectrum is the superposition of the spectra
of several different active blobs, all producing (in equilibrium)
roughly the same slope.

Different models of the formation of active blobs can share
the general features outlined above. Following GRV we discuss
a specific model below.

\vskip 1 truecm
\centerline{\bf 3. STRUCTURED CORONA: A SIMPLE MODEL}
\vskip 0.5 truecm
{\it 3.1 Time scales}
\vskip 0.5 true cm
\noindent
It has been realized that the magnetic field can be amplified by
differential rotation in accretion disks to equipartition
values and can then provide the needed viscosity and energy dissipation
mechanisms.
A discussion of the different ways in which this could occur can be found in
e.g. the review of Hayvaerts 1990.
Here we show that by adopting the simple model of GRV we can
derive plausible values for the parameters introduced phenomenologically
in the previous section, i.e. the ratio $ t_d/  t_c$ and the size $R_b$
of the blobs, as well as the number of active blobs at any time.

In the dynamo model of GRV, the disk differential rotation causes the
azimuthal magnetic field $B_{\phi}$ to grow exponentially because of a
feedback mechanism linking the radial and vertical components of the
magnetic field {\bf B} to $B_{\phi}$.
The magnetic fields is amplified until its pressure equals the
surrounding gas pressure, which leads to buoyancy of the magnetic
flux tubes (see Stella \& Rosner 1984).
While reconnection of the magnetic field lines {\it within} the disk is
ineffective in preventing the growth of the magnetic field, it probably
occurs very rapidly in the much more tenuous coronal medium,
transforming the stored energy in kinetic energy of fast particles.
Within the disk, $B_{\phi}$ grows as $\dot B_{\phi}=(3
v/R)B_{\phi}$, where the convection velocity $v$ is $\approx \alpha^{1/3}
\Omega z_0$.
Here $\alpha$ is the parameter linking the stress tensor to the total
disk pressure, $\Omega=\sqrt{GM/R^3}$ is the Keplerian angular
velocity, and $z_0$ is the disk scale height.
We can then identify the charge time scale $ t_c$ with $R/3v$.

{}From the standard theory of accretion disks (Shakura \& Sunyaev 1973), in
the radiation pressure dominated region
$z_0 \approx 9 R_s {\cal L} S(r)$
where ${\cal L}$ is the total luminosity of the source in Eddington
luminosity units, and $S(r)=1-\sqrt{3/r}$.
Here $R_s$ is the Schwartzchild radius $R_s\equiv 2GM/c^2$, and
$r\equiv R/R_s$.
According to GRV, the size of the loops is $R_b \approx z_0/\alpha^{1/3}$.
Note that, since $z_0 \propto {\cal L}$, the compactness
$\ell$ of a single loop ($\propto  L/R_b$) is independent
of the source luminosity.

With the above estimates the charge time scale becomes
$$
t_c \sim
{R \over 3\alpha^{1/3}\Omega z_0}\, \sim \,
0.5{r^{5/2}\, M_6 \over \alpha^{1/3} {\cal L}\,S(r)} \qquad
{\rm s},
\eqno(4)
$$
where $M_6\equiv M/10^6M_{\odot}$.
Note that in  sources closer to the Eddington limit the magnetic field
grows rapidly because of the higher convection velocity
$(v\propto z_0 \propto {\cal L})$.

The discharge time scale $ t_d$ depends on the microphysical
processes that cause the dissipation of magnetic energy.
We parameterize it as
$$
t_d\, =\, a\left({R_b\over c} \right).
\eqno(5)
$$
$t_d$ is the maximum between the time needed to transfer the magnetic
energy to the particles and the cooling time scale of the particles.
Since the cooling time is very short (see below), the strongest
constraint derives from the acceleration time scale which is highly
uncertain, but should be $\sim R_b/v_{rec}$, where $v_{rec}\,$ is
the reconnection velocity. Theoretical estimates of $v_{rec}\,$ range from
$v_{rec}\,\sim
v_A/\ln R_m$, where $v_A$ is the Alfv\`en velocity and $R_m$ the
magnetic Reynolds number for  Petschek's type reconnection (Petschek 1964),
to $v_{rec}\sim v_A$ (Priest \& Forbes 1986).

A rough estimate of $a$ can be derived from the $\sim 10$ sec duration
of the impulsive
phase of Solar flares during which particles are believed to
be accelerated.
Typical solar values of $R_b$ and $v_A$ are $\sim 10^9$ cm and $\sim
6\times 10^8$ cm/s respectively, so that the acceleration time is $\sim
6$ times the Alfv\`en wave crossing time.
Assuming that in the AGN rarefied coronal medium $v_A \sim c$,
we obtain $a\sim 6$.
The ratio $\rat$ can then be written from equations (4) and (5) as
$$
{ t_c \over  t_d}\sim 3\, \left({r\over 5}\right)^{5/2}\,
\left({10 \over a}\right)\, \left({0.1 \over {\cal L}}\right)^2\,
\left[{1\over S(r)}\right]^2,
\eqno(6)
$$
independent of viscosity and mass.
At $r=5$ we obtain $\rat \simgt 60$ (for ${\cal L}\simlt 0.1$)
which ensures that $Q_{rep}\gg Q_{disk}$ throughout the disk.
The ratio $\rat$ becomes of order unity in the innermost part of the
disk ($5\simlt r\simlt 12$) only for sources radiating close to the
Eddington limit.

Since the cooling time of the accelerated particles is shorter than the
time scale of the energy release in the loops (see below) we can
think of the spectral evolution as a succession of stationary states.
The energy input in the hot particles may fluctuate strongly on the
time scale associated with the evolution of magnetic structures in the
loop phase (minutes-hours for Seyferts), while keeping a constant
average on medium time scales (days), determined by the constancy in the
energy transport of the accretion flow.

Using the time scale estimates given above, the loop variability time scale
should be of the order of minutes for a $10^6$ solar mass object, and of
milliseconds for a 10 solar mass object (see also Pudritz 1981a,b, Pudritz \&
Fahlman 1982, and Abramowicz et al. 1991,  for a detailed discussion of
variability in corona models).

\vskip 0.7 true cm
{\it 3.2 Number and luminosity of active loops}
\vskip 0.5 true cm

\noindent
We assume that the dynamo process operates all over the disk. In a
disk sector between $R-R_b$ and $R+R_b$ the number of magnetic loops that
are growing {\it within} the disk is $N_{grow}=4R/R_b$, and the
number of active loops {\it above} the disk is $N_{act}=( t_d/  t_c)N_{grow}$.
Transforming these two quantities into differentials, [i.e. $dN/dr =
N/(2R_b)$], and substituting the values of $\rat$ and $R_b$ derived
above, we obtain the differential number of active loops.
Integrating from $3R_s$ to infinity, we then derive the total number
of active loops at any time, that is  $N_{tot}\sim 5  a \alpha^{2/3}$.
This number does not depend neither on the luminosity nor on the mass of
the source.

According to Pringle (1981), observations of stellar accretion disks
seem to require $\alpha \sim 0.1$.
Setting $a=10$ we find $N_{tot}\sim 10$.
{}From simple Poissonian noise arguments, this is what is needed to explain
X--ray fluctuations of a factor $\sim 2$ on short time scales, that are
typically observed in AGN and GBH.

Finally the luminosity of a single blob located at $r$ is
estimated by
$$
\eqalign{
L_{blob}\, & = \,f Q(R)\, 2\pi R\, {dR \over dN_{act}} \cr
{}~& \sim \,
1.2\times 10^{43}\, M_6\, \left({f\over 0.2}\right) \,
\left({10\over a}\right) \, \left( {0.1\over \alpha} \right)^{2/3}\,
\left( {1\over r} \right)^{1/2}\,{\cal L}\, S(r)
\quad {\rm erg\, s^{-1}}.}
\eqno(7)
$$
Since $S(r)=1-\sqrt{3/r}$,
the most luminous blobs are those located at $r=12$ (although the
maximum of the coronal surface emissivity is at $r\simeq 5$) and can have
30\% of the total luminosity emitted by the entire ensemble of loops.
{}From the above equation, the compactness parameter of the loops is
$$
\ell \, \sim \,
50\,
\left({f\over 0.2}\right) \, \left({10\over a}\right) \,
\left( {0.1\over \alpha} \right)^{1/3}\,
\left( {1\over r} \right)^{1/2}.
\eqno(8)
$$
In the most radiative part of the disk the blobs have $\ell \sim 30$.
This should be considered as a {\it lower limit} of the actual
compactness, since the acceleration region may be smaller than
$R_b$, the value used in the derivation of equation (8).
The compactness $\ell$ is independent of the source luminosity and the mass
of the central black hole. This may account for the remarkable
similarity of the high energy spectra of AGN and GBH.
Furthermore, the order of magnitude of the compactness parameter is
consistent with observations (Done \& Fabian 1990).
In retrospect, since the Compton cooling time in a source of compactness $\ell$
is $\sim (R/c)/\ell$ our assumption that the cooling time is shorter
than the acceleration time is justified.

\vskip 0.7 true cm
{\it 3.3 Reprocessed radiation}
\vskip 0.5 true cm
\noindent
The temperature of the thermal radiation below the loops will
be generally higher than the temperature of the disk emission at the
same radius producing a hotter thermal component superimposed to the
multicolor disk emission.
The flux per unit time per unit area of the reprocessed radiation below
the loops is a factor
$\rat$ higher than that  due to viscous dissipation in the disk.
Using a black body approximation, the temperatures
of the two components can be written as
$$
kT_{disk}\, \sim \,
30 \, M_6^{-1/4}\,(1-f)^{1/4} \left({5\over r}\right)^{3/4}
\, \left({{\cal L}\over 0.1}\right)^{1/4}\, S(r)^{1/4}\ \ \ \
{\rm eV}, \eqno(9.a)
$$
$$
kT_{rep}\, \sim \, 35 \, M_6^{-1/4}\, f^{1/4}\,\left({10\over a}\right)^{1/4}\,
\left({5\over r}\right)^{1/8}\,
\left({0.1\over {\cal L}}\right)^{1/4} S(r)^{-1/4}\ \ \ \ {\rm eV}. \eqno(9.b)
$$
In the innermost part of the accretion disk, the reprocessed component
can be $\approx 2-3$ times hotter than the disk temperature at the same
radius.
The reprocessed radiation temperature is a very weak function of the
radius and is higher in low luminosity objects.
The total thermal emission from the acreting flow is then formed by a
multicolor spectrum as in the standard $\alpha$--disk (Shakura \&
Sunyaev 1973) --with the difference that at each $R$ the surface
emissivity is a fraction $(1-f)$ of that of a standard disk-- plus a
hotter multicolor component with a much smoother radial dependence.

\vskip 1 true cm
\centerline{\bf 4. DISCUSSION}
\vskip 0.5 true cm

We have proposed that the formation of a structured corona above an
accretion disk can be responsible for the X--ray emission of AGN and GBH.
The patchy structure of the corona can be due to the formation of
magnetic loops in which the energy is stored over a long `charge' time
and is subsequently released (by, e.g. reconnection) over a much shorter
`discharge' time.
Each blob can then produce X--rays in the same way as the
homogeneous corona discussed in Papers I and II but, at the
same time, the present model can account for a broader set of observational
constraints.

The model includes three emission components:
1) the luminosity due to dissipation
within the accretion disk which we generally call $L_{UV}$ (although in
GBH it is mostly emitted in the soft X--ray band); 2) the upward
Comptonized luminosity  $L_X$ extending from the medium to the hard X-ray
range; 3) the reprocessed luminosity $L_{rep}$ which comprises a thermal peak
hotter than the direct disk emission and a Compton reflection hump at
10--100 keV (Lightman \& White 1988).

The ratio of $L_X$ to $L_{UV}$ is determined by the fraction $f$ of the
total gravitational power stored in the magnetic field,
while  $L_{rep}$ is {\it always} of the order of $L_X$.
Observations of Seyfert galaxies
suggest that the UV emission may be up to an order of
magnitude larger than the X--ray emission (e.g. Walter \& Fink 1993):
in the present model this corresponds to $f=0.2$.
For $f=0.2$ the high energy spectrum can have a spectral
index close to unity if $\rat \gg 5$, which is not a severe constraint.

The reprocessed component $L_{rep}$
may be related to the variable hard UV component
inferred in some Seyfert galaxies (e.g. NGC 4151) and/or to the so
called ``soft-excess" observed in the soft X--ray band (e.g. Yaqoob \& Warwick
1991, Pounds et al. 1993, Walter et al. 1993).
The soft excess luminosity relative to the the direct
disk emission $L_{UV}$ is determined by $f$.

The actual size scale and number of active regions and the associated
variability time scales, estimated in \S 3
on the basis of the GRV model for the
magnetic corona, are generally consistent with observations.
In particular the number of blobs turns out to be $\sim 10$,
independent of the luminosity of the source and of the mass of the
central black hole.
Short time scale variability can be associated
with stochastic noise in the number of active blobs which can
easily produce flux variability up to a factor 2.
Since the size of the blobs scales with total luminosity $L$, the variability
time scale turns out to be proportional to $L$.

$L_X$ can fluctuate on the shortest timescales, and its fluctuation is
associated with the blob size and the energy dissipation rate.
$L_{rep}$
must vary in a correlated fashion with  $ L_X$.
However a smearing
in time can result from the shape of the blobs which may illuminate a larger
area than that intercepted by the blobs themselves ($C<1$; \S 2).
In addition there is a  dilution introduced by the disk component $L_{UV}$.
In objects where the UV component is
definitely more luminous than the X--ray component ($f\ll 1$)
the dilution is expected to be important.
In these cases we do not expect correlated variability on short time scales
between the UV and the X--rays.
However, the correlation could hold on medium time scales
if the power dissipated in the disk varies and $f$ is constant.
On the other hand, we predict a substantial amount of UV flux correlated to
the X--rays in sources where these two components have comparable luminosity.
We note that in the two cases where
the UV -- X-ray correlation has been observed,
NGC 5548 (C92) and  NGC 4151 (Perola et al. 1986, 1994), the UV and X--ray
fluxes are indeed comparable. It is therefore possible,
pending an estimate of the  bolometric correction
factors for the two bands (e.g. Perola and Piro 1994), that
for these objects $f$ is close to unity
and therefore the reprocessed component is strong in the UV.
The lack of short time variations in UV (C92)
suggests that the reprocessing region is greater than the size of the blobs.
This is consistent with a moderate photon starvness of the
emitting blobs (i.e. $C<1$), as required by the flatness of
the [2--20] keV spectrum of these two objects [$\alpha_x \simeq 0.5$ for NGC
4151 (Yaqoob \& Warwick 1991) and $\alpha_x \simeq 0.8$ for NGC 5548 (Nandra
et al. 1991)].

We note that further reprocessing of the emitted radiation can occur
in other regions further away from the central
black hole, like the intermediate
accretion disk, the broad line clouds and the obscuring torus. A discussion
of such mechanisms is beyond the scope of this Letter.

\vskip 1 true cm
\centerline{\bf 5. SUMMARY}
\vskip 0.5 true cm

Our results can be summarized as follows:

\item{i)} In the limit $f\gg  t_d/ t_c$
the ratio between the Comptonized luminosity and the luminosity injected
in soft photons in each blob is independent of $f$, i.e.
the fraction of gravitational power carried by the magnetic field.
This ensures that the spectral index of the Comptonized component is
largely independent on details and different blobs produce
spectra with the same power law index.
The ratio $L_{UV}/L_X$ can be large in the present model, since not
all the accretion power is assumed to be released in the hot corona
(cf. Papers I and II).

\item{ii)}The number and compactness
of active blobs are independent of the source
luminosity and of the central black hole mass.

\item{iii)} The actual value of the X--ray spectral index is controlled
by the geometry of the blobs.
Quasi--spherical blobs will be photon starved producing flatter
spectra and a smearing of the reprocessed component in space and time.

\item{iv)} Short--term variability of the X--ray luminosity
can be associated with
the stochastic noise in the number and luminosity of active blobs.
A reprocessed component with temperature higher than the disk emission
and with luminosity similar to the X-ray luminosity is predicted
to follow the X--ray variations with some smearing on shortest timescales.

\vskip 1 true cm
\noindent
\centerline{\bf ACKNOWLEDGEMENTS}
\vskip 0.5 true cm
We thank the second anonimous referee for greatly improving
the style and the clarity of the manuscript.


\vskip 1 true cm
\noindent
\centerline{\bf REFERENCES}
\vskip 0.5 true cm

\parindent=0 pt
\everypar={\hangindent=2.6pc}

Abramowicz, M.A., Bao, G., Lanza, A., \& Zhang, X.-H. 1990, A\&A, 245, 454

Cameron, R.A., et al. 1993, Proceedings of Compton Symposium, ed N. Gehrels,
St. Luis, in press

Clavel, J., et al. 1992, ApJ, 393, 113 (C92)

Done, C., \& Fabian, A.C. 1990, MNRAS, 240, 81

Done C., Mulchaey, J.S., Mushotzky, R.F., \& Arnaud, K.A. 1992, ApJ, 395, 275

Galeev, A.A, Rosner, R., \& Vaiana, G.S. 1979, ApJ, 229, 318 (GRV)


Grebenev, S., et al. 1993, A\&AS, 97, 281

Haardt, F., Done, C., Matt, G., \& Fabian, A.C. 1993, ApJ, 411, L95

Haardt, F., \& Maraschi, L. 1991, ApJ, 380, L51 (Paper I)

Haardt, F., \& Maraschi, L. 1993, ApJ, 413, 507 (Paper II)

Heyvaerts, J. 1990, Proceedings of 6th IAP
Astrophysical Meeting / IAU coll., ed. Bertout et al., Edition Frontieres,
p129

Johnson, W.N., et al. 1993, A\&AS, 97, 21


Lightman, A.P., \& White, T.R. 1988, ApJ, 335, 57

Maisack, M., et al. 1993, ApJ, 407, L61

Nandra, K., et al. 1991, MNRAS, 248, 760

Perola, G.C., et al. 1986, ApJ, 306, 508

Perola, G.C., \& Piro, L. 1994, A\&A, 281, 7

Petschek H.E., 1964,  AAS--NASA Symposium on the physics of solar flares,
NASA Special Publ. No. 50, p.425

Pounds, K.A., Nandra, K., Fink, H.H., \& Makino, F. 1994, MNRAS, 267, 193

Pozdnyakov, L.~A., Sobol', I.~M., \& Sunyaev, R.~A. 1983,
Ap. Space Sc. Rev., Vol. 2, p. 189

Priest, E.R., \& Forbes T.G. 1986, J.Geophys.Res., 91, 5579

Pringle, J.E. 1981, ARA\&R, 19, 137

Pudritz, R.E. 1981a, MNRAS, 195, 881

Pudritz, R.E. 1981b, MNRAS, 195, 897

Pudritz, R.E., \& Fahlman, G.G. 1982, MNRAS, 198, 689

Shakura, N.I., \& Sunyaev, R.A. 1973, A\&A, 24, 337

Shapiro, S.L., Lightman, A.P., \& Eardly, D.M. 1976, ApJ, 204, 187

Stella, L., \& Rosner, R. 1984, ApJ, 277, 312


Sunyaev, R.A., \& Titarchuk, L.G. 1980, A\&A, 86, 121

Ueda, Y., Ebisawa, K., \& Done, C. 1994, PASJ, 46, 107

Walter, R., \& Fink, H.H. 1993, A\&A, 274, 105

Walter, R., Orr, A., Courvoisier, T.J.--L., Fink, H.H., Makino, F., Otani, C.,
\& Wamsteker, W. 1994, A\&A, submitted


\bye